\documentclass[12pt]{article}
\pdfoutput=1
\usepackage{amsmath,amssymb,amsthm,bm,graphicx,float,array,multirow,multicol,rotfloat,caption,subcaption,hyperref,cleveref,enumerate,geometry,mathdots,adjustbox,booktabs,parskip,mathtools,tikz,tikz-cd,pdflscape,csquotes,lscape,rotating,empheq,mathrsfs,longtable}
\usepackage[para]{threeparttable}
\usepackage[all]{xy}
\usepackage[normalem]{ulem}
\usepackage[aligntableaux=center]{ytableau}
\usepackage[numbers,sort&compress]{natbib}
\numberwithin{equation}{section}
\numberwithin{figure}{section}
\allowdisplaybreaks
\makeatletter
\usetikzlibrary{arrows,shapes.misc,positioning,decorations.pathmorphing,decorations.markings,decorations.pathreplacing,matrix,patterns,backgrounds}
\newcommand\scalemath[2]{\scalebox{#1}{\mbox{\ensuremath{\displaystyle #2}}}}

\definecolor{gold}{RGB}{255,215,0}
\definecolor{purple}{RGB}{160,32,240}
\tikzset{
scale cd/.style={every label/.append style={scale=#1}, cells={nodes={scale=#1}}},
gauge/.style={rounded rectangle, draw=black!100, thick, minimum size=5mm}, 
gaugeD/.style={rounded rectangle, draw=black!100,double,thick,minimum size=5mm},  
empty/.style={rounded rectangle, draw=white!100, thick, minimum size=5mm}, 
flavor/.style={rectangle, draw=black!100, thick, minimum size=5mm},
flavorD/.style={rectangle, draw=black!100, double,thick, minimum size=5mm},
node/.style={circle, thick, draw=black!100,fill=white!100,  minimum size=2mm, inner sep=0pt},
sqnode/.style={rectangle
, thick, draw=black!100,fill=white!100,  minimum size=2mm, inner sep=0pt
},
sonode/.style={circle, thick, draw=black!100,fill=red!100,  minimum size=3mm, inner sep=0pt},
spnode/.style={circle, thick, draw=black!100,fill=blue!100,  minimum size=3mm, inner sep=0pt},
fnode/.style={rectangle, thick, draw=black!100,fill=white!100,  minimum size=3mm, inner sep=0pt},
tnode/.style={rounded rectangle, outer sep=0pt, thick, minimum size=5mm},
brace/.style={decoration={brace, mirror},decorate}
}

\newcommand\notsoscript{\@setfontsize\notsoscript{9}{7}}

\theoremstyle{plain}
\newtheorem*{thm*}{Theorem}

\theoremstyle{definition}

\newtheorem*{defn*}{Definition}

\makeatother

\def\scheme/{our affine scheme} 

\graphicspath{{figures/}}

\begin{document}

\begin{titlepage}
\vspace*{-3cm} 
\begin{flushright}
{\tt DESY-25-105}\\
\end{flushright}
\begin{center}
\vspace{1.2cm}
{\LARGE\bfseries Index from a point}
\vspace{1.2cm}

{\large
Monica Jinwoo Kang,$^{1,2}$ Craig Lawrie,$^{3}$ and Jaewon Song$^{4}$\\}
\vspace{.7cm}
{$^1$ Department of Physics and Astronomy, University of Pennsylvania\\
Philadelphia, PA 19104, U.S.A.}\par
\vspace{.2cm}
{$^2$ Mitchell Institute for Fundamental Physics and Astronomy, Texas A\&M University,\\
College Station, TX 77843, U.S.A.}\par
\vspace{.2cm}
{$^3$ Deutsches Elektronen-Synchrotron DESY,\\
Notkestr.~85, 22607 Hamburg, Germany}\par
\vspace{.2cm}
{$^4$ Department of Physics, Korea Advanced Institute of Science and Technology\\
Daejeon 34141, Republic of Korea}\par
\vspace{.2cm}

\vspace{.3cm}

\scalebox{.9}{\tt monicak@tamu.edu, craig.lawrie1729@gmail.com, jaewon.song@kaist.ac.kr}\par
\vspace{1.2cm}
\textbf{Abstract}
\end{center}
{
We propose an algebro-geometric interpretation of the Schur and Macdonald indices of four-dimensional $\mathcal{N}=2$ superconformal field theories (SCFTs). We conjecture that there exists an affine scheme $X$, which reduces to the Higgs branch as a variety, such that the Hilbert series of the (appropriately-graded) arc space of its polynomial ring $J_\infty(\mathbb{C}[X])$ encodes the indices. 
Distinct local descriptions of a (singular) point correspond to distinct choices of $X$, giving rise to families of $\mathcal{N}=2$ SCFTs each without a Higgs branch. These local descriptions directly translate into nilpotency relations in the operator product expansions. We test our conjecture across a variety of (generalized) Argyres--Douglas theories. 
} 
\vfill 
\end{titlepage}

\tableofcontents
\newpage

\section{Introduction}

One of the key insights gained over the past few decades in quantum field theory (QFT) is that the conventional Lagrangian and path-integral formulation does not capture the full landscape of consistent quantum field theories. This realization stems from the existence of a wide variety of inherently strongly-coupled theories, often referred to as ``non-Lagrangian'' QFTs, which defy a standard Lagrangian description. In the conventional approach, a QFT is defined by specifying a set of fields and their interactions encoded in a Lagrangian. However, such a framework proves inadequate for describing these exotic non-Lagrangian theories. This raises a natural and pressing question: is there a more general formulation of QFT that can encompass both conventional and non-Lagrangian theories on equal footing?

The conformal bootstrap program offers an alternative approach to study QFTs. It attempts to solve conformal field theories (CFTs) by imposing constraints arising from symmetry and consistency (such as unitarity), yielding universal constraints on the spectrum and correlation functions for a generic CFT with the imposed symmetry property. As it imposes powerful, model-independent bounds that apply to any theory sharing the same symmetry structure, it may not produce a sharp prediction for a particular CFT. Remarkably, in certain cases, such as the Ising model and the $O(N)$ model, the bootstrap has led to precise determinations of the spectrum and correlators. A common feature in these successes is the emergence of a ``decoupling'' in the space of CFT data, thereby severely constraining the spectral properties and dynamics. For instance, in the BPZ minimal models in 2d \cite{Belavin:1984vu} or the 3d critical Ising model \cite{El-Showk:2014dwa}, there is a truncation in the spectrum of primary operators. Similar phenomena have also been observed to play a crucial role in 4d $\mathcal{N}=2$ SCFTs \cite{Liendo:2015ofa, Lemos:2015orc}.

Motivated by these observations, we explore whether the decoupling phenomenon can be elevated to a defining characteristic of a specific CFT. In this paper, we conjecture that for 4d $\mathcal{N}=2$ SCFTs, particular limits of the supersymmetric partition function (namely, the Schur or Macdonald index) can be entirely constructed by a decoupling or vanishing condition in the spectrum. We propose a translation of this decoupling condition into algebraic geometry in terms of nilpotency relations, such as $x^n = 0$.\footnote{This idea is motivated by the behavior of operator product expansions (OPEs), where certain components vanish when the same primary operator is multiplied several times. Such a phenomenon is realized, for example, in the $(A_1, A_{2n})$ Argyres--Douglas theory \cite{Agarwal:2018zqi}, where the short multiplet $\widehat{\mathcal{C}}_{n(\frac{n}{2}, \frac{n}{2})}$, which generically appears in the OPE of the form $(\widehat{\mathcal{C}}_{0(0, 0)})^{n+1}$, is absent.} While the equation $x^n = 0$ naively describes the point $x = 0$, viewing it in the context of polynomial rings reveals the richer structure imposed by the nilpotent relation. This distinction reflects the essential difference between a scheme and a variety in algebraic geometry, and it plays a central role in our proposal.

Let $\mathcal{T}$ denote some 4d $\mathcal{N}=2$ SCFT. We propose that there exists an affine algebraic scheme $X$, necessarily bifiltered to capture the structure of the superconformal algebra, associated with $\mathcal{T}$ such that the underlying variety of $X$ is the Higgs branch of $\mathcal{T}$:
\begin{equation}
    X_\text{red} = \mathcal{M}_{H}(\mathcal{T}) \,.
\end{equation}
Furthermore, $X$ encodes the indices of $\mathcal{T}$ in the following way. Given any scheme $Y$, there exists a construction of an auxiliary space (another affine scheme), called the arc space, which we denote as $J_\infty (Y)$. We will also study the ring of polynomials over $Y$ and $J_\infty(Y)$:
\begin{align}
   Y \leadsto  J_\infty(Y) \ ,\qquad R=\mathbb{C}[Y] \leadsto J_\infty(R) = \mathbb{C}[J_\infty(Y)] \,.
\end{align}
As the name suggests, the arc space $J_\infty(Y)$ encodes all local paths (arcs) passing through the space $Y$. This allows us to further study the local or singular properties of $Y$. We review the definitions and properties of the arc space (and the, related, jet schemes) in Section \ref{sec:arc}. Given a grading on $Y$, which we refer to via the $q$-degree, the arc space has a natural bigrading, and one can determine the Hilbert series of the (polynomial ring over the) arc space with respect to this bigrading. One obtains a polynomial:
\begin{equation}
    h_Y(p, q) = \operatorname{HS}_{p, q}(J_\infty(\mathbb{C}[Y])) \,.
\end{equation}
Here, the fugacity $q$ denotes the grading of the monomials that exists before going to the arc space, and $p$ denotes the arc space grading. One can further refine the Hilbert series by additional, ``flavor'', symmetries, if they are present in $\mathcal{T}$. 

In this paper, we propose that the Schur index \cite{Gadde:2011ik, Gadde:2011uv} of the 4d $\mathcal{N}=2$ SCFT $\mathcal{T}$, which is a limit of the full superconformal index \cite{Kinney:2005ej, Romelsberger:2005eg}, can be recovered from the Hilbert series of the arc space of an appropriately chosen affine scheme $X$:
\begin{align}\label{eqn:schurconj}
\boxed{
    I_S(q) = h_X(q, q) = \operatorname{HS}_{p, q}(J_\infty(\mathbb{C}[X])) \Big|_{p\to q}
} \,.
\end{align}
The Schur index counts 1/4-BPS states of the theory, and it is invariant under continuous marginal deformations. We emphasize that we have to choose a particular relative grading prescription for $p$ and $q$ to match the Hilbert series with the index of a 4d theory. 

More generally, the $q$-grading of $X$ extends to part of a bifiltration structure on $X$, where the second part of the bifiltration captures the $\mathfrak{su}(2)_R$ symmetry of the $\mathcal{N}=2$ superconformal algebra. Keeping track of the full bifiltration, we can then construct an associated graded ring $\operatorname{gr}(J_\infty (R))$ to $J_\infty(R) = J_\infty(\mathbb{C}[X])$, which is trigraded. The Hilbert series is then a polynomial in three fugacities, $q$, $p$, and $T$, for which a certain specialization provides the Macdonald index \cite{Gadde:2011uv} of $\mathcal{T}$:
\begin{align}\label{eqn:macconj}
\boxed{
    I_{\textrm{Mac}}(q, T) = \operatorname{HS}_{p,q,T}\left(\mathrm{gr}(J_\infty (R)) \right) \Big|_{p\to q}
} \,.
\end{align}
The $T \rightarrow 1$ limit, which recovers the Schur index, corresponds to throwing up the information about the full bifiltration and keeping only the $q$-grading. We will explain this expression in detail in Section \ref{sec:mac}, and we also refer to \cite{Kang:2026nge}.  This is essentially a geometric analogue of the procedure in \cite{Song:2016yfd}. The Macdonald index counts the same set of states as the Schur index, but with refined information of $\mathfrak{su}(2)_R$ grading. 

While it may be tempting to claim that all affine schemes give rise to (the indices of) a 4d theory, we will see later that not all affine schemes have arc spaces that can be associated to indices of consistent 4d $\mathcal{N}=2$ SCFTs. Instead, we might conjecture that there exists a restricted class of affine schemes, specified by currently unknown algebro-geometric conditions,\footnote{In \cite{Kang:2026nge}, the algebro-geometric conditions are worked out for the $(A_{k-1}, A_{N-1})$ Argyres--Douglas theories with $\gcd(k,N)=1$.} such that for each such affine scheme there is a corresponding 4d $\mathcal{N}=2$ superconformal field theory $\mathcal{T}_X$ modulo exactly marginal deformations (and also possible global structures):\footnote{Here the choice of global structures, such as the 1-form symmetry or discrete theta terms may not be specified. Therefore, at best it maps to a relative theory.}
\begin{align}\label{eq:bolcconj}
 \{ \textrm{Affine schemes}' \} &\to \{ \textrm{4d } \mathcal{N}=2 ~\textrm{(relative) SCFTs (up to marginal deformations)} \} \,.
\end{align}
However, it is unclear whether even the full superconformal index, which is still far from the full spectrum of the theory, can be entirely determined from this data. Nevertheless, we do not even have any examples of distinct theories with identical Schur index,\footnote{There are cases where the Schur indices of two distinct theories are identical up to a rescaling of fugacities \cite{Buican:2019kba, Buican:2020moo, Kang:2021lic}.} and so such a conjecture might not be as bold as it first appears.

Concretely, in this paper, we find the following correspondence between affine schemes and 4d $\mathcal{N}=2$ SCFTs of Argyres--Douglas type \cite{Argyres:1995jj, Argyres:1995xn, Eguchi:1996vu, Xie:2012hs}:
\begin{align}
\begin{array}{c|c}
    \mathcal{T}_X & X \\
    \hline
    (A_1, A_{2n}) & x^{n+1} = 0 \\
    (A_2, A_3) & x^2 + y^3 = 0 \ ,\quad xy=0 \\
    (A_2, A_4) & x^2 + y^3 = 0\ ,\quad xy^2=0 \\
    (A_2, A_6) & x^3 + 4xy^3 = 0\ ,\quad 3x^2y + y^4 = 0 \\
    (A_2, A_7) & y^5 + 5y^2x^2 = 0\ ,\quad 2x^3 + 3xy^3 = 0 \\
    (A_3, A_4) & 4xz + y^2 + 8z^3 = 0, \quad xy + 4yz^2 = 0, \quad x^2 + 4y^2z + 12xz^2 +  22z^4 = 0 \\
    (A_1, D_{2n+1}) & w = x^2 + y^2 + z^2 \ , \quad xw^n = yw^n = zw^n=0 \\
    (A_1, A_{5}) & xy+z^3 - t z = 0 \, \quad xt = 0 \,, \quad yt =0 \,, \quad z(xy+z^3) = 0
\end{array}
\end{align}
Notice that each of the spaces $X$ above (except for the last two entries) are as affine varieties simply a point. However, they are all different as affine schemes. 
The correspondence between the arc space of a point for the $(A_1, A_{2n})$ theory has already been found in \cite{Bhargava:2023hsc}. 

Our correspondence is motivated by the following reasoning. 
Let $V_\mathcal{T}$ denote the vertex operator algebra associated to $\mathcal{T}$ by the correspondence of \cite{Beem:2013sza}. 
The Schur index of $\mathcal{T}$ is identical to the vacuum character of the associated VOA:
\begin{equation}
    I_S(q) = \chi_{V_\mathcal{T}}(q) \,.
\end{equation}
The VOA captures the Schur sector of the 4d theory. The Schur sector includes the Higgs branch chiral ring, and it is indeed possible to recover the Higgs branch moduli space from the VOA $V_\mathcal{T}$. To this end, we first consider Zhu's $C_2$-algebra \cite{zhu1996modular}, which is defined as
\begin{equation}
    R_{V} = V / C_2(V) \,, 
\end{equation}
where 
\begin{align}
    C_2(V) = \{ a^{(1)}_{-h_1-i_1} a^{(2)}_{-h_2-i_2} \cdots|0 \rangle : i_m \ge 1 \} \,. 
\end{align}
Here $a^{(1)}, a^{(2)}, \cdots$ are the strong generators of $V$ and $h_m$ are the weights of the primary fields. To put it simply, Zhu's $C_2$-algebra $R_V$ is the commutative counterpart of $V$, dropping all the higher descendants coming from the derivatives; in particular, the $C_2$-algebra captures certain decoupling relations in the OPEs of operators belonging to the Schur sector of $\mathcal{T}$.
Now, the associated scheme is defined as \cite{MR3456698}
\begin{equation}
    X_V = \operatorname{Spec}(R_V) \,,
\end{equation}
and the algebraic variety underlying the associated scheme, called the associated variety, can be obtained by reducing the scheme, dropping the nilpotents.
This is conjectured to be isomorphic to the Higgs branch of the 4d SCFT \cite{Beem:2017ooy}:
\begin{equation}
    (X_V)_\text{red} = \mathcal{M}_H(\mathcal{T}) \,.
\end{equation}
Since the spectrum of $R_V$ encodes decoupling relations in the OPEs of certain protected operators, and is conjectured to reduce as a variety to the Higgs branch, we are motivated to ask if there exists a ring $R$ with $X = \operatorname{Spec}(R)$, which might or might not be $X_V$, that captures all decoupling relations necessary to reconstruct the index.\footnote{We note that if the VOA is \emph{classically-free} \cite{MR4779563}, then the conjecture in equation \eqref{eqn:schurconj} holds for $X = X_V$. It has been conjectured (see \cite{RastelliTalk1,RastelliTalk2}) that all VOAs that are associated to 4d $\mathcal{N}=2$ SCFTs are classically-free. We thank Chris Beem and Leonardo Rastelli for discussions on this point.}

In general, it is not expected that the information of the Higgs branch $(X_V)_\text{red}$ can be used to reconstruct the Schur sector of a 4d $\mathcal{N}=2$ theory. This is because the Schur sector contains contributions from multiplets that are opaque to the Higgs branch chiral ring. However, based on the conjecture in equation \eqref{eqn:schurconj}, it appears that keeping the nilpotents and considering the unreduced scheme instead of the Higgs branch does allow us to reconstruct (at least the index of) the Schur sector.

In fact, in this paper, we will largely focus on examples of 4d $\mathcal{N}=2$ SCFTs, $\mathcal{T}$, where the Higgs branch chiral ring does not contain \emph{any} operators; the Higgs branch is a point:
\begin{equation}
    \mathcal{M}_H(\mathcal{T}) = (X_V)_\text{red} = \operatorname{Spec}(\mathbb{C}) = \{ \operatorname{pt} \} \,.
\end{equation}
Such examples are drawn from the $(A_{k-1}, A_{N-1})$ Argyres--Douglas SCFTs with $\gcd(k, N)=1$.
For example, Zhu's $C_2$-algebra for the $(A_1, A_{2n})$ Argyres--Douglas theory is given as \cite{Beem:2017ooy}
\begin{align}
    (A_1, A_{2n}): R_V = \mathbb{C}[x]/( x^{n+1} ) \,.
\end{align}
This gives a trivial associated variety $(X_V)_\text{red}$, consistent with the absence of the Higgs branch. However, the associated scheme $X_V$ is non-trivial. One can now compute the Hilbert space of the arc space $J_\infty (R_V)$, which gives the Schur/Macdonald index upon appropriate mapping of fugacities. 
This is already demonstrated in \cite{Bhargava:2023hsc} (and mathematically proved for $(A_1, A_2)$ in \cite{bai2020quadratic}), and we will further generalize it to higher-rank Argyres--Douglas theories and also the ones with a non-trivial Higgs branch. For the higher-rank Argyres--Douglas theories, the computation of the Macdonald index requires the construction of the graded algebra $\text{gr}(J_\infty(R))$, whereas for the $(A_1, A_{2n})$ theory, this was unnecessary. 

Our goal is to ask the opposite direction. For a given affine scheme $X$, when do we have a 4d $\mathcal{N}=2$ superconformal theory whose Hilbert series of $J_\infty(\mathbb{C}[X])$ reproduces the index? We start with various descriptions of singularity, identical as an affine variety but different as an affine scheme. 
We find that whenever we obtain a valid index of a 4d SCFT, in most cases, our affine scheme $X$ is identical to the associated scheme $X_V$, which is identical to the \emph{non-reduced} spectrum of the Zhu's $C_2$-algebra $R_V$ of the associated VOA $V$ \cite{Beem:2013sza, Beem:2017ooy}. However, there is a subtlety here: our affine scheme comes equipped with a bifiltration, and while $X$ is isomorphic to $X_V$ as schemes, they may not be isomorphic as \emph{bifiltered} schemes. One important open question is to understand the corresponding bifiltration structure on $X_V$. This is especially prominent for theories with non-trivial Higgs branch, which we discuss in Section \ref{sec:higgsable}, where we find it important to add to the ring $R$ an additional generator, beyond the strong generators of the VOA, and interpreted as corresponding to the stress-tensor.

The structure of this paper is as follows. In Section \ref{sec:higgsschemes}, we discuss the arc space and our computational method. We also review the nilpotency structure of the Argyres--Douglas theories. In Section \ref{sec:nonhiggsable}, we consider various (singular) affine schemes where the underlying variety is a point, and find that they give rise to Argyres--Douglas theories without a Higgs branch. In Section \ref{sec:mac}, we describe the method to compute the Macdonald index, and explicitly determine the index of the $(A_2, A_3)$ Argyres--Douglas theory. In Section \ref{sec:higgsable}, we consider affine schemes that reduce to the simplest Kleinian singularities and find that they give rise to Argyres--Douglas theories with non-trivial Higgs branches. Finally, we conclude in Section \ref{sec:conc} with some future directions.

\emph{Note added:} When finalizing the current article, a very interesting preprint \cite{Andrews:2025krn} appeared on arXiv, which has some overlapping results with ours. The current paper mostly focuses on the case with (higher-rank) Argyres--Douglas theories with a trivial Higgs branch. The results of this paper were initially presented by one of the authors (J.S.) at Seoul National University on May 23rd, 2025. 

\section{Arc spaces and nilpotency relations of AD theories} \label{sec:higgsschemes}

\subsection{Arc spaces: a primer}\label{sec:arc}

We propose in this paper that the Schur and Macdonald indices of a four-dimensional $\mathcal{N}=2$ SCFT can be obtained from the arc space of a certain affine scheme. In this section, we give a short practical preliminary to the notion of arc spaces (or arc schemes), providing sufficient information for computing the Hilbert series. For further background on jet schemes and arc spaces, we refer the reader to \cite{MR2483946,MR2820711}. A detailed discussion of the relationship between arc spaces and vertex algebras can be found in \cite{ArakawaMoreauBook}.

In the context of relevance to this paper, we consider the arc space associated to an affine scheme of the form
\begin{equation}
    X = \operatorname{Spec} \left( \mathbb{C}[x_1, \cdots, x_m] /  ( f_1, \cdots, f_r ) \right) \,,
\end{equation}
where each $f_i$ in the ideal is a homogeneous polynomial in the coordinates $x_1, \cdots, x_m$, with respect to a set of assigned degrees $d_j = \operatorname{deg}(x_j)$. These degrees are constrained by the relations $f_i$, though they are generally not uniquely fixed by the homogeneity of the relations in the ideal alone.
The $n$th jet scheme, denoted $J_n(X)$, is constructed by promoting each coordinate to a formal power series
\begin{equation}
    x_j \,\,\to\,\, x_j(t) = \sum_{\alpha = 0}^n x_j^{(\alpha)} t^\alpha \,,
\end{equation}
and expanding the defining relations modulo $t^{n+1}$. That is, for each $f_i$, we write
\begin{equation}
   f_i \mod t^{n+1} \quad =\,\, \sum_{\alpha=0}^n f_i^{(\alpha)} t^{\alpha} \,.
\end{equation}
Then, the $n$th jet scheme $J_n(X)$ is given by
\begin{equation}
    J_n(X) = \operatorname{Spec} \left( \mathbb{C}[x_1^{(0)}, \cdots, x_1^{(n)}, \cdots, x_m^{(0)}, \cdots, x_m^{(n)}] \, / \, F  \right) \,,
\end{equation}
where the ideal $F$ is the ideal generated by the coefficients of the expansions,
\begin{equation}
    F =  ( f_1^{(0)}, \cdots, f_1^{(n)}, \cdots, f_r^{(0)}, \cdots, f_r^{(n)} ) \,.
\end{equation}
The jet scheme $J_n(X)$ admits a natural bigrading defined by
\begin{equation}
    \operatorname{deg}(x_j^{(\alpha)}) = (d_j, \alpha) \,,
\end{equation}
where $d_j$ is the original degree of $x_j$, and $\alpha$ corresponds to the order of the derivative.
The arc space of $X$, denoted $J_\infty(X)$, is defined as the inverse limit of this tower of jet schemes:
\begin{equation}
    J_\infty(X) = \varprojlim_{n} J_n(X) \,.
\end{equation}
Practically, this involves enumerating the infinitely many coordinates and polynomials in the ideal for the arc space of a given scheme $X$.
Intuitively, the arc space captures formal smooth paths (or arcs) based at points of $X$, with the bidegree keeping track of derivative order. For example, the first jet scheme $J_1(X)$ can be identified with the total space of the tangent bundle of $X$.

With this explicit description of the arc scheme $J_\infty(X)$, we can define and compute the Hilbert series of the associated polynomial ring,
\begin{equation}
    J_\infty(R) = \mathbb{C}[J_\infty(X)] \,,
\end{equation}
where $R=\mathbb{C}[X]$ is the polynomial ring of the original scheme $X$. In general, the ideal relations $f_i^{(\alpha)}$ admit nontrivial syzygies, and computing the Hilbert series typically requires first finding a Gr\"obner basis. This computation has been carried out analytically for certain tractable cases, such as the arc space of $\operatorname{Spec}(\mathbb{C}[x]/(x^n))$ in the unrefined (single-parameter) case \cite{MR2820711}, and for the fully refined Hilbert series in the case $n=2$ in \cite{bai2020quadratic}.

To compute the Hilbert series of the arc space, it is often sufficient to truncate to a finite level and instead compute the Hilbert series of the $n$th jet scheme. This gives the Hilbert series of the arc space up to an $n$-dependent order in the expansion. Specifically, the contributions to the Hilbert series from the coordinates and relations in the ideal begin at the following orders:
\begin{equation}
    x_j^{(\alpha)} \, : \quad p^{\,\operatorname{deg}(x_j)}\, q^\alpha \,, \qquad f_i^{(\alpha)} \, : \quad p^{\,\operatorname{deg}(f_i)} \, q^\alpha \,.
\end{equation}
Here, the degree $\operatorname{deg}(x_j)$ assigned to each variable may be constrained by the relations $f_i$, but the overall normalization often remains ambiguous. In physical applications, this ambiguity is typically resolved by input from four-dimensional field theory, and different choices may correspond to different observables. We will elaborate on this point in Section \ref{subsec:AD}.

Given this setup, it becomes clear that to compute the Hilbert series $\operatorname{HS}_{p,q}(J_\infty(R))$ up to order $\mathcal{O}(q^n)$,\footnote{That is, including the term of order $q^n$.} it suffices to compute $\operatorname{HS}_{p,q}(J_n(R))$. In particular, for the case relevant to our conjecture for the Schur index, where we take $p = q$ in the Hilbert series,\footnote{We refer to this specialization as the Schur limit of the Hilbert series.}, we aim to compute
\begin{equation}
    \operatorname{HS}_{q, q}(J_\infty(R)) \,,
\end{equation}
up to a given order in $q$ by computing the Hilbert series of the corresponding jet scheme.

Let $d$ denote the smallest degree among the variables $x_j$. Then, to compute the Schur limit of the Hilbert series up to and including order $q^\ell$, it suffices to compute
\begin{equation}
    \operatorname{HS}_{q, q}(J_\infty(R)) \,\,=\,\, \operatorname{HS}_{q, q}(J_n(R)) \qquad \text{ mod } \qquad q^{d + n + 1} \,,
\end{equation}
which implies that
\begin{equation}
\operatorname{HS}_{q, q}(J_\infty(R)) \quad \text{to order } q^\ell \quad \text{is determined by} \quad \operatorname{HS}_{q, q}(J_{\ell - d}(R)) \,.
\end{equation}
The computation of the Hilbert series for a jet scheme generally requires determining a Gr\"obner basis for the defining ideal. This can be efficiently done using a software package for commutative algebra. In our work, we use {\tt Macaulay2} \cite{M2} for these computations.

\subsection{Vanishing short multiplets in Argyres--Douglas theories} \label{subsec:AD}

In this subsection, we review some of the key facts about the Argyres--Douglas (AD) theories that are relevant for the remainder of the paper. Argyres–Douglas theories are 4d $\mathcal{N}=2$ superconformal field theories that generically have Coulomb branch operators with non-integer scaling dimensions \cite{Argyres:1995jj, Argyres:1995xn, Eguchi:1996vu, Xie:2012hs, Wang:2015mra}. As a consequence, they do not admit conventional Lagrangian descriptions with manifest $\mathcal{N}=2$ supersymmetry. These theories typically arise as infrared fixed points of $\mathcal{N}=2$ supersymmetric field theories where various mutually non-local electromagnetically charged particles become simultaneously massless.\footnote{Some AD theories can also be realized as fixed points of renormalization group flows from $\mathcal{N}=1$ Lagrangian theories \cite{Maruyoshi:2016tqk, Maruyoshi:2016aim, Agarwal:2016pjo, Agarwal:2017roi, Benvenuti:2017bpg}, providing a powerful framework to study their properties using supersymmetric Lagrangian tools.}

One motivation for studying Argyres–Douglas theories is that they can be regarded as among the ``simplest'' examples of four-dimensional $\mathcal{N}=2$ SCFTs. This simplicity is reflected in their associated vertex operator algebras (VOAs): for instance, the $(A_1, A_{2n})$ theories correspond to Virasoro minimal models, the $(A_{k-1}, A_{n-1})$ theories (with $\gcd(k,n) = 1$) correspond to $\mathcal{W}$-algebra minimal models \cite{Cordova:2015nma, Buican:2015ina}, and the $D_p(G)$ theories (with $\gcd(p, h_G^\vee) = 1$) \cite{Cecotti:2012jx, Cecotti:2013lda} correspond to affine vertex algebra or simple affine Lie algebra \cite{Xie:2016evu}. Notably, the $(A_1, A_2)$ theory saturates the minimal value of the central charge $c$ \cite{Liendo:2015ofa}, and its associated VOA is the Yang--Lee model, the simplest Virasoro minimal model.

These correspondences are further supported by explicit computations of the Schur index for these theories \cite{Buican:2015ina, Buican:2015tda, Song:2015wta, Buican:2017uka, Song:2017oew}, which exhibit remarkably concise closed-form expressions. In addition, the BPS spectrum on the Coulomb branch of these theories is particularly tractable \cite{Shapere:1999xr}, making them ideal laboratories for exploring the intricate connections between BPS states in the Coulomb branch, CFT operator spectra, and vertex algebras \cite{Cecotti:2010fi, Iqbal:2012xm, Cordova:2015nma, Cecotti:2015lab, Cordova:2017mhb, Gaiotto:2024ioj, Kim:2024dxu}.

Argyres--Douglas theories also admit a variety of string and M-theoretic realizations \cite{Shapere:1999xr, Gaiotto:2009hg, Xie:2012hs, Wang:2015mra}. In particular, the Type IIB construction provides a remarkably simple description for the so-called $(G, G')$ theories, where $G$ and $G'$ are ADE Lie algebras. These theories arise from Type IIB string theory on a Calabi--Yau threefold singularity given by a hypersurface of the form
\begin{align}
    W(x, y, z, w) = W_G (x, y) + W_{G'} (z, w) = 0 \,, \quad (x, y, z, w) \in \mathbb{C}^4 ,
\end{align}
where $W_G$ and $W_{G'}$ are the defining polynomials of ADE surface singularities, given as
\begin{subequations}
\begin{align}
    W_{A_n} (x, y) &= x^{n+1} + y^2 \,,\\
    W_{D_n} (x, y) &= x^{n-1} + x y^2\,, \\
    W_{E_6} (x, y) &= x^3 + y^4 \,,\\
    W_{E_7} (x, y) &= x^3 + x y^3 \,, \\
    W_{E_8} (x, y) &= x^3 + y^5 \,.
\end{align}
\end{subequations}
The resulting low-energy effective theory in 4d is a $(G, G')$ theory, which is an $\mathcal{N}=2$ SCFT. 
Deformations of the complex structure of the singularity correspond to parameters in the 4d theory, including the Coulomb branch operators. We will focus on the conformal phase of the theory, which is at the singularity above.

A characteristic feature of Argyres–Douglas theories is the absence of certain short multiplets that would generically appear in the operator product expansion (OPE) of universal operators such as the stress tensor or flavor currents \cite{Liendo:2015ofa, Agarwal:2018zqi}. For instance, in the $(A_1, A_{2n})$ theory, the OPE of stress tensor multiplets of the form $T^{n+1} = (\widehat{\mathcal{C}}_{0(0,0)})^{n+1}$ would typically contain the short multiplet $\widehat{\mathcal{C}}_{n(\frac{n}{2}, \frac{n}{2})}$.\footnote{We follow the notation of \cite{Dolan:2002zh}.} However, this multiplet is absent in the $(A_1, A_{2n})$ theory, either because it is not part of the spectrum or because the corresponding OPE coefficient vanishes.

This phenomenon can be understood from the perspective of the associated VOA. The VOA corresponding to the $(A_1, A_{2n})$ theory is the Virasoro minimal model $M(2, 2n+3)$, which contains a null state of the form $((L_{-2})^{n+1} + \cdots)\Omega$ at level $2n$, where $\Omega$ denotes the vacuum state. This reflects the vanishing of the $\widehat{\mathcal{C}}_{n(\frac{n}{2}, \frac{n}{2})}$ multiplet in the physical theory.

This vanishing condition is also captured by the Schur index of the theory, which takes the following form \cite{andrews1999a2, Cordova:2015nma, Song:2015wta, Song:2017oew}:
\begin{align}\label{eqn:dora}
I_{(A_1, A_{2n})}(q) = \mathrm{PE}\left[ \frac{q^2 - q^{2n+2}}{(1 - q)(1 - q^{2n+3})} \right]\,,
\end{align}
where $\operatorname{PE}$ denotes the standard plethystic exponential. 
The term $\frac{q^2}{1 - q}$ arises from the contribution of the stress tensor, while the term $- \frac{q^{2n+2}}{1-q}$ reflects the absence of the $\widehat{\mathcal{C}}_{n(\frac{n}{2}, \frac{n}{2})}$ multiplet.\footnote{The Schur index alone does not uniquely determine which short multiplet is absent; this requires additional input from OPE selection rules \cite{Liendo:2015ofa, Ramirez:2016lyk, Kiyoshige:2018wol, Gimenez-Grau:2020jrx}. In some cases, further information can be obtained by studying the Macdonald or full superconformal index, combined with character decomposition of short multiplets \cite{Agarwal:2018zqi, Song:2021dhu}.}

Motivated by this observation, we define the affine scheme $X$ corresponding to the $(A_1, A_{2n})$ theory via the vanishing relation:
\begin{align}
(A_1, A_{2n}): \quad x^{n+1} = 0 \,.
\end{align}
Thus, the associated scheme is $X = \operatorname{Spec} \mathbb{C}[x]/(x^{n+1})$, which is known as a ``fat point".\footnote{Notably, this coincides with the associated scheme \cite{MR3456698}, defined as the spectrum of Zhu’s $C_2$-algebra: $\operatorname{Spec}(R_V)$ with $R_V = \mathbb{C}[x]/(x^{n+1})$ \cite{Beem:2017ooy}.}
As we will show in Section \ref{sec:nonhiggsable}, this single relation $T^{n+1} = 0$, or equivalently the affine scheme $X$, is sufficient to determine the entire Schur index, and even the Macdonald index, of the theory without requiring any additional constraints. This is particularly striking given that $T^{n+1} = 0$ represents only one of the infinitely many vanishing short multiplets in the spectrum!

Now, we consider some examples of higher-rank Argyres--Douglas theories. The associated VOA for $(A_{k-1}, A_{N-1})$ is given by the $W(k, k+N)$ minimal model. For the $k=3$ and $N=3m+r$ ($r=1, 2$) case, we have two strong generators $T$ and $W$ whose spins are $2$ and $3$, and there are null states of the form $((W_{-3})^{m+1} + \cdots)\Omega$ and $ ((W_{-3})^m(L_{-2})^{r} + \cdots )\Omega$, which can be understood as a phenomenon of vanishing short multiplets in a certain OPE \cite{Agarwal:2018zqi}. This is consistent with the known Schur index \cite{andrews1999a2, Song:2017oew}, which is given as
\begin{align}
    I_{(A_2, A_{N-1})}(q) = \textrm{PE} \left[ \frac{q^2 + q^3 - q^{N+1} - q^{N+2}}{(1-q)(1-q^{3+N})} \right] \,. 
\end{align}
For low values of $N$, this motivates us to define the affine schemes as follows:
\begin{subequations}
\begin{align}
    (A_2, A_3): &\quad x^2 + y^3 = 0 \,,\ xy=0 \,,\\
    (A_2, A_4): &\quad x^2 + y^3 = 0\,,\ xy^2=0 \,.
\end{align}
\end{subequations}
Here $x$ and $y$ comes from $W$ and $T$ respectively. Importantly, as there are few monomials that can contribute at the relevant degrees, any coefficients can be absorbed by linear redefinitions of the coordinates. Thus, for $N = 4, 5$, we do not need to concern ourselves with the coefficients; for $N > 5$, this is no longer the case. For $N = 7, 8$, the coefficients appearing in the polynomials at each degree can be fixed (up to linear redefinitions) by demanding that the Hilbert series of the arc space reproduces the Schur index. We find 
\begin{subequations}
\begin{align}
    (A_2, A_6): &\quad x^3 + 4xy^3 = 0 \,,\ 3x^2y + y^4 = 0 \,,\\
    (A_2, A_7): &\quad y^5 + 5y^2x^2 = 0\,,\ 2x^3 + 3xy^3 = 0 \,.
\end{align}
\end{subequations}
In each of these cases, the scheme reduces to a point when reduced to an affine variety. But, as schemes, they have a more intricate singularity structure, which captures the physics of the indices.

We can further generalize this to an arbitrary $(A_{k-1}, A_{N-1})$ theory. The associated VOA is a $W_k$-algebra, containing $k-1$ generators of spins $2, 3, \cdots, k$. The Schur index (or the vacuum character of the $W_k$-algebra) is given by
\begin{align}\label{eqn:schurAA}
    I_{(A_{k-1}, A_{N-1})}(q) = \textrm{PE} \left[ \frac{q^2+q^3+\cdots +q^k-q^{N+1}-q^{N+2}-\cdots - q^{N+k-1}}{(1-q)(1-q^{k+N})} \right] \,. 
\end{align}
One can read off generators and relations from the index above. For example, consider the $(A_3, A_4)$ theory. The numerator of the index inside the PE reads $q^2 + q^3 + q^4 - q^6 - q^7 - q^8$. The first three terms give the generators, say, $W_2, W_3, W_4$, and the last three terms give the relation. The relations that are consistent with the index are
\begin{align}
\begin{split}
    q^6: &\quad W_4 W_2 + (W_3)^2 + (W_2)^3 = xz + y^2 + z^3 = 0 \,,\\
    q^7: &\quad W_4 W_3 + W_3 (W_2)^2 = xy+yz^2= 0 \,, \\
    q^8: &\quad (W_4)^2 + (W_3)^2 W_2 + (W_2)^4 + (W_2)^2 W_4 =x^2 + y^2 z + z^4 + z^2 x= 0 \,,
\end{split}
\end{align}
where we ignored the coefficients. Writing $(W_2, W_3, W_4) \to (z, y, x)$ will give the desired relations for the point-like singularity in three dimensions. It is possible to get the same relations from $F=z^5 + z^3 x + z x^2 + z^2 y^2 + y^2 x$ and taking derivatives with respect to $x, y, z$.\footnote{ 
More generally, it is conjectured \cite{Xie:2019zlb} that for the Argyres--Douglas theories of type $(A_{k-1}, A_{n-1})$ with $(k, n)=1$, Zhu's $C_2$-algebra $R_V$ is given by a simple Jacobi algebra of the form 
\begin{align*}
 R_V = \mathbb{C}[w_1, \ldots, w_\ell] \Big/ \left(\frac{\partial F}{\partial w_1}, \ldots \frac{\partial F}{\partial w_\ell} \right) \,, 
\end{align*}
where $F=F(w_1, \ldots w_\ell)$ is the sum of monomials of fixed degree and the degrees for $w_i$ are given as $i+1$.} 
However, the numerical coefficients appearing in the polynomials that define the affine scheme affect the Hilbert series of the arc space, and therefore it is necessary to be precise about the coefficients before attempting to match the indices. In general, the Jacobi algebra which is conjectured to be identical with Zhu's $C_2$-algebra does not produce the correct Schur index.

\section{Points: non-Higgsable Argyres--Douglas theories} \label{sec:nonhiggsable}

In this section, we explicitly compute the Hilbert series of the arc space associated with various ``schemifications'' of a point. Remarkably, we find that these reproduce the Schur indices of Argyres–Douglas theories that lack a Higgs branch. As we will demonstrate, distinct algebraic descriptions of a single ``point'' can give rise to a rich multitude of distinct 4d theories.

\subsection{\texorpdfstring{$(A_1, A_2)$}{(A1, A2)} theory}

Let us compute the Hilbert series of the arc space of the double point $X = \textrm{Spec}(R)$ with $R = \mathbb{C}[x]/( x^2 )$ in a brute-force manner. The exact formula has already been obtained (and proved) in \cite{bai2020quadratic}. Here, we perform the computation order-by-order for illustration purposes. 
In this case, it turns out that we do not need to introduce the associated graded vector space to obtain the Macdonald index. 

The arc space for this double point is described by expanding the coordinate $x$ and thus the relation $x^2$ generating the ideal as formal power series:
\begin{align}
    x(t) = \sum_{\alpha=0}^\infty x^{(\alpha)} t^\alpha \,, \quad x(t)^2  = \sum_{\alpha=0}^\infty f^{(\alpha)}(x^{(\alpha)}) t^\alpha = 0 \,. 
\end{align}
Now, assign bidegree to $x^{(\alpha)}$ as $\operatorname{deg}(x^{(\alpha)}) = (2, \alpha)$; then, one can see that $\operatorname{deg}(f^{(\alpha)}) = (4, \alpha)$. The arc space of the double point is given as 
\begin{align}
    J_\infty (X) = \operatorname{Spec}\left(\mathbb{C}[x^{(0)}, x^{(1)}, \cdots]/( f^{(0)}, f^{(1)}, \cdots ) \right) \,. 
\end{align}
To compute the Hilbert series, notice that the arc space of the polynomial ring over $X$ can be decomposed according to this bidegree to get 
\begin{equation}\label{eqn:kirby}
    J_\infty(R) = \mathbb{C}[x^{(0)}, x^{(1)}, \cdots]/( f^{(0)}, f^{(1)}, \cdots )  = \bigoplus_{n, m=0}^{\infty} V_{n, m} \,,
\end{equation}
where the Hilbert series of the polynomial ring over $J_\infty(X)$ is thus given as
\begin{align}
    \operatorname{HS}_{p, q}(J_\infty(R)) = \sum_{n, m=0}^{\infty} \textrm{dim}V_{n, m} \, q^n p^{2m} \ . 
\end{align}
Here $V_{n, m}$ can be thought of as the vector space spanned by $m$ letters of $x_i$, whose subscripts add up to $n$ modulo relations. 

First, we consider the spaces spanned by single letters in $x^{(\alpha)}$: $V_{n, 1} = \langle x^{(n)} \rangle$. This gives rise to a $q^n p^2$ term in the Hilbert series. Summing over $n$, we find 
\begin{align}
    \sum_{n=0}^\infty \operatorname{dim} V_{n, 1} \, q^n p^2 = \frac{p^2}{1-q} \,.
\end{align}
Next, at order $p^4$, we have 
\begin{equation}
    V_{n, 2} = \langle x^{(\alpha)} x^{(n-\alpha)}: \alpha=0, \cdots \lfloor n/2 \rfloor \rangle \,/ \, ( f^{(n)} ) \,.
\end{equation}
Now, at each order in $n$, the number of generators is identical to the number of partitions of $n$ into elements $1$ and $2$. The generating function of such partitions is given as $1/(1-q)(1-q^2)$. 
At order $q^n p^4$, we also notice that one relation, $f^{(n)}$, kills one of the generators, so that we get the contribution of $-q^n p^4$. Summing up in $n$, we get
\begin{align}
    \sum_{n=0}^\infty \operatorname{dim} V_{n, 2} \, q^n p^4 = \frac{p^4}{(1-q)(1-q^2)} - \frac{p^4}{1-q} = \frac{p^4 q^2}{(q;q)_2} \ , 
\end{align}
where $(q; q)_n \equiv \prod_{i=1}^n (1-q^i) $ is the $q$-Pochhammer symbol. We see the pattern that the first term in the middle counts the number of generators and the second term with the minus sign counts the number of relations at given order. 

We can proceed to order $p^6$. At this level, the generator gives contribution of $p^3/(1-q)(1-q^2)(1-q^3)$. The relations are trickier to count, since there is a relation among the relations that affects the counting. At order $q^n p^6$, the relevant relations are of the form $x^{(\alpha)} f^{(n-\alpha)}$, with $\alpha=0, 1, \cdots n$, giving $n+1$ of them. In addition to this, we have the \emph{first syzygy}, which is a relation among the relations, at this order given as
\begin{align}
    \sum_{\alpha=0}^n (x^{(n-\alpha)} f^{(\alpha)})(2n-3\alpha) = 0 \,, 
\end{align}
which reduces the number of relations by one, except for $n=0$. Thus, the generating function for the relations, including the effect of syzygies, is
\begin{align}
    1 + \sum_{n=1}^\infty n q^n = 1 + \frac{q}{(1-q)^2} = \frac{1+q^3}{(1-q)(1-q)^2} \,.
\end{align}
Therefore, the Hilbert series of the arc space at order $p^6$ is given as
\begin{align}
    \sum_{n=0}^\infty \operatorname{dim} V_{n, 3} \, q^n p^6 = \frac{p^6}{(1-q)(1-q^2)(1-q^3)} - \frac{p^6 (1+q^3)}{(1-q)(1-q^2)} = \frac{p^6 q^6}{(q;q)_3} \,. 
\end{align}

One can proceed in such a manner to prove that the exact formula for the Hilbert series for the arc space of the double point, $J_\infty (X)$, is given as \cite{bai2020quadratic}
\begin{align}
    \operatorname{HS}_{p,q}(J_\infty(R)) = \sum_{n, m \ge 0} \frac{q^n p^{2n(n-1)}}{(q;q)_n} \ . 
\end{align}
Now, if we take $p^2=q^2 T$, we see that we obtain the fermionic or Nahm-sum expression of the Macdonald index of $(A_1, A_2)$ theory \cite{Foda:2019guo}, which up on taking $T \to 1$ reduces to the Schur index or the vacuum character of the Virasoro minimal model $M(2, 5)$.

Note that if we take $p^2=q$, we instead obtain the non-vacuum character of the $M(2, 5)$ minimal model. This is identical to the Schur index of the $(A_1, A_2)$ theory with a surface defect insertion \cite{Cordova:2017mhb}. Instead, this can be thought of as the vacuum super-character of $\mathfrak{osp}(1|2)_1$ whose central charge is $c=2/5$ \cite{Creutzig:2024ljv}.

\subsection{\texorpdfstring{$(A_1, A_4)$}{(A1, A4)} theory}

A similar analysis to that of $(A_1, A_2)$ can be performed for $(A_1, A_4)$, where our affine scheme is simply
\begin{equation}
    X = \operatorname{Spec} \left( \mathbb{C}[x] / ( x^3 ) \right) \,,
\end{equation}
and where the degree of the single coordinate $x$ is $\operatorname{deg}(x) = 2$. Again one can write the decomposition of the arc space according to the bigrading as we did for $(A_1, A_2)$ in equation \eqref{eqn:kirby}. Then, we simply need to compute the dimensions of the vector spaces $V_{n,m}$ taking into account the relations and syzygys. Instead of proceeding by hand, we can ask {\tt Macaulay2} to determine a Gr\"obner basis and perform the calculation for jet schemes of sufficiently low jet degree. As discussed already, we know that 
\begin{equation}
    \operatorname{HS}_{p,q}(J_n(R)) = \operatorname{HS}_{p,q}(J_\infty(R)) \quad \mod q^{n+1} \,.
\end{equation}

For example, by computing the Hilbert series of the polynomial ring of the $8$th jet scheme, we find the Hilbert series of the polynomial ring of the arc space, to leading orders in $q$:
\begin{equation}
\begin{aligned}
    \operatorname{HS}_{p,q}(J_\infty(R)) &= 1+p^2+p^4 + (p^2+p^4) q + (p^2+ 2p^4+p^6) q^2 + (p^2+2p^4+2p^6) q^3\\
    &\quad + (p^2+3p^4+3p^6+p^8) q^4+ (p^2+3p^4+4p^6+p^8) q^5\\
    &\quad  + (p^2+4p^4+6p^6++3p^8) q^6 + (p^2+4p^4+7p^6+4p^8) q^7 \\
    &\quad + (p^2+5p^4+9p^6+7p^8+p^{10}) q^8 + \cdots \,.
\end{aligned}
\end{equation}
Since our affine scheme is defined as the spectrum of a polynomial ring in a single variable, it is not necessary to construct the associated graded algebra $\operatorname{gr}(J_\infty(R))$ to find the Macdonald index; the Macdonald limit if simply $p^2 \rightarrow q^2 T$. We can see that this produces the known expression \cite{Song:2015wta, Song:2016yfd, Foda:2019guo} for the Macdonald index of the $(A_1, A_4)$ Argyres--Douglas theory. Looking at the Schur limit of the Hilbert series, $p = q$ or $T = 1$ in the Macdonald limit, we can see that
\begin{equation}
    \operatorname{HS}_{q,q}(J_\infty(R)) = \sum_{i=0}^{16} a_i q^i + \mathcal{O}(q^{17}) \,,
\end{equation}
where the coefficients $a_i$ are
\begin{equation}
    (1, 0, 1, 1, 2, 2, 3, 3, 5, 6, 8, 9, 13, 14, 19, 22, 28) \,.
\end{equation}
It is straightforward to see that these coefficients are reproduced by the formula for the Schur index, given in equation \eqref{eqn:dora}, for the $(A_1, A_4)$ Argyres--Douglas theory.

Let us note that if we take $p^2=q$ limit, we find
\begin{align}
   1+q+2 q^2+2 q^3+3 q^4+4 q^5+6 q^6+7 q^7+10 q^8+12 q^9+ O(q^{10}) \,, 
\end{align}
which agrees with the character of the non-vacuum module with weight $h_{1, 3}=-3/7$ for the $M(2, 7)$ minimal model. 

\subsection{\texorpdfstring{$(A_1, A_6)$}{(A1, A6)} theory}

It is straightforward to perform the same analysis for $(A_1, A_6)$ exactly as we did for $(A_1, A_4)$, and thus we do not excessively belabor the point here. Our affine scheme is 
\begin{equation}
    X = \operatorname{Spec} \left( \mathbb{C}[x] / ( x^4 ) \right) \,,
\end{equation}
where $\operatorname{deg}(x) = 2$, as usual. We can again compute the Hilbert series of the polynomial ring of the $8$th jet scheme to find that
\begin{equation}
\scalemath{.94}{
\begin{aligned}
    \operatorname{HS}_{p,q}(J_\infty(R)) &= 1+p^2+p^4+p^6 + \left( p^2+p^4+p^6 \right) q + \left( p^2 + 2p^4 + 2p^6 + p^8 \right)q^2 \\
    &\ + \left( p^2 + 2p^4 + 3p^6 + 2p^8\right) q^3 + \left(p^{2} + 3p^4 + 4p^6 + 4p^8  + p^{10}\right)q^4 \\
    &\ + \left(p^{2} + 3p^4 + 5p^6 + 5p^8 + 2p^{10}\right)q^5 + \left(p^{2} + 4p^{4} + 7p^6 + 8p^8 + 4p^{10} + p^{12}\right)q^6\\
    &\ + \left(p^{2}  + 4p^4+ 8p^6 + 10p^8 + 5p^{10} + p^{12}\right)q^7 + \left(p^2 + 5p^4 + 10p^6 + 14p^8 \right.\\
    &\quad\; \left.  + 10p^{10} + 2p^{12}\right)q^8 + \cdots \,.
\end{aligned}}
\end{equation}
The $p^2 = q^2 T$ limit reproduces the known Macdonald index of the $(A_1, A_6)$ theory (see Appendix A.1 of \cite{Song:2016yfd}) to $q^{10}$ order, and the $T = 1$ limit reproduces the Schur index, as given in equation \eqref{eqn:dora}. The first coefficients in the Schur limit of the Hilbert series are
\begin{equation}
    (1, 0, 1, 1, 2, 2, 4, 4, 6, 7, 10, 12, 17, 19, 26, 31, 40, 47) \,,
\end{equation}
as determined via truncation to the $15$th jet scheme. In these $(A_1, A_{2n})$ examples, it is straightforward to go to much higher orders numerically using {\tt Macaulay2}; while we have verified that these are consistent with our proposal, we do not write the full expressions here for brevity.

\subsection{\texorpdfstring{$(A_2, A_3)$}{(A2, A3)} theory}

As an indicative example, we explore the conjecture in equation \eqref{eqn:schurconj} in the context of the $(A_2, A_3)$ Argyres--Douglas theory in explicit detail. As determined in Section \ref{subsec:AD}, \scheme/ is
\begin{equation}
    X = \operatorname{Spec}(\mathbb{C}[x, y] / (x^2 + y^3, xy)) \,,
\end{equation}
where the variables are graded as
\begin{equation}
    \operatorname{deg}(x) = 3\,, \quad \operatorname{deg}(y) = 2 \,,
\end{equation}
whereby we can see that each equation in the ideal is homogeneous. The $n$th jet scheme is 
\begin{equation}
    J_n(X) = \operatorname{Spec}(\mathbb{C}[x_0, y_0, \cdots, x_n, y_n] / I_n) \,,
\end{equation}
where $I_n$ is the ideal generated by the $2n+2$ polynomials,
\begin{equation}
  \begin{aligned}
    \sum_{\substack{i, j \geq 0 \\[0.2em] i + j = m}} x_i x_j - \sum_{\substack{i, j, k \geq 0 \\[0.2em] i + j + k = m}} y_i y_j y_k\ ,\qquad \sum_{\substack{i, j \geq 0 \\[0.2em] i + j = m}} x_i y_j \qquad & \left( m = 0, \cdots, n\right) \,.
  \end{aligned}
\end{equation}
Recalling that $x_i$ has bidegree $(3, i)$ and $y_i$ has bidegree $(2, i)$, we can now determine the bigraded Hilbert series for $J_n(R)$ by brute-force. We find that
\begin{equation}
  \begin{aligned}
    \operatorname{HS}_{p, q}(J_{n \geq 4}(R)) &= 
    (1 + p^2 + p^3 + p^4 + p^6) + 
    (p^2 + p^3 + p^4 + p^5 + p^6)q \\&\quad + 
    (p^2 + p^3 + 2 p^4 + 2 p^5 + 3 p^6 + p^7 + p^8)q^2 \\&\quad +
    (p^2 + p^3 + 2 p^4 + 3 p^5 + 4 p^6 + 2 p^7 + 2 p^8 + p^9)q^3 \\&\quad +
    (p^2 + p^3 + 3 p^4 + 4 p^5 + 6 p^6 + 4 p^7 + 5 p^8 + 2 p^9 + p^{10})q^4 + \cdots
    \,.
  \end{aligned}
\end{equation}
The terms that we have written explicitly are fixed for all $n \geq 4$, however the higher-order terms in the $\cdots$ of course depend on the specific choice of $n$. Specializing to the Hilbert series of the arc space where the fugacities are identified, we find:\footnote{To determine the coefficients written explicitly here, it was only necessary to determine the Hilbert series of the $12$th jet scheme, as discussed.}
\begin{equation}
  \begin{aligned}
    \operatorname{HS}_{q, q}(J_\infty(R)) = 1 &+ q^2 + 2q^3 + 3q^4 + 3q^5 + 6q^6 + 7q^7 + 11q^8 + 14q^9 + 20q^{10} \\ 
    &+ 25q^{11} + 36q^{12} + 44q^{13} + 60q^{14} + \cdots  \,.
  \end{aligned}
\end{equation}
It is easy to see that this matches the expansion of the Schur index for the $(A_2, A_3)$ theory as given in equation \eqref{eqn:schurAA}.

\subsection{\texorpdfstring{$(A_2, A_4)$}{(A2, A4)} theory}

For the $(A_2, A_4)$ Argyres--Douglas theory, \scheme/ is
\begin{equation}
    X = \operatorname{Spec}(\mathbb{C}[x, y] / (x^2 + y^3, xy^2)) \,,
\end{equation}
where the variables are graded as
\begin{equation}\label{eqn:A2A4grading}
    \operatorname{deg}(x) = 3\,, \quad \operatorname{deg}(y) = 2 \,.
\end{equation}
It is now straightforward to use {\tt Macaulay2} to determine the Hilbert series of the associated arc space. We find
\begin{equation}
    \operatorname{HS}(J_\infty(R)(q,q) =  \sum_{j = 0}^{15} a_j q^j + \mathcal{O}(q^{16}) \,,
\end{equation}
where the coefficients $a_j$ are given by
\begin{equation}
    (1, 0, 1, 2, 3, 4, 7, 8, 14, 18, 26, 34, 49, 62, 86, 112) \,.
\end{equation}
To determine these coefficients, we truncated the arc space of the $13$th jet scheme, $J_{13}(R)$, which, as discussed, has the same coefficients in the Schur limit of the Hilbert series as the arc space up to and including order $q^{15}$. The Schur index for the $(A_2, A_4)$ theory was given in equation \eqref{eqn:schurAA}, and expanding that expression we can see that it agrees with the Schur limit of the Hilbert series as given here.

\subsection{\texorpdfstring{$(A_2, A_6)$}{(A2, A6)} theory}

The analysis for the $(A_2, A_6)$ theory follows in largely the same vein as the $(A_2, A_4)$ theory. Our affine scheme was mentioned in Section \ref{sec:higgsschemes}; the degrees of the polynomials were fixed by the known orders of the null relations in the associated W-algebra, and the coefficients are fixed by requiring the reproduction of the Schur index when computing the Hilbert series of the arc space. In this way, we find the affine scheme
\begin{equation}
    X = \operatorname{Spec}(\mathbb{C}[x, y] / (x(x^2 + 4y^3), y(3x^2 + y^3)) \,.
\end{equation}
The variables $x$ and $y$ are graded in the same way as for $(A_2, A_4)$ as given in equation \eqref{eqn:A2A4grading}. Now we can write down the coordinates and ideal defining the arc space, and we can compute the Schur limit of the Hilbert series to a particular order by truncating to the appropriate jet scheme. We find that 
\begin{equation}
    \operatorname{HS}(J_\infty(R))(q,q) =  \sum_{j = 0}^{16} a_j q^j + \mathcal{O}(q^{17}) \,,
\end{equation}
where the coefficients $a_j$ are given by 
\begin{equation}
    (1, 0, 1, 2, 3, 4, 8, 10, 16, 22, 33, 44, 66, 86, 121, 162, 221) \,.
\end{equation}
The coefficients are again determined via an explicit computation of the Hilbert series of the $14$th jet scheme using {\tt Macaulay2}. These again match with the expansion of the Schur index for the $(A_2, A_6)$ theory as given in equation \eqref{eqn:schurAA}.

\subsection{\texorpdfstring{$(A_2, A_7)$}{(A2, A7)} theory}

In the final explicit example of the $(A_2, A_{N-1})$ sequence of Argyres--Douglas SCFTs, we consider the $(A_2, A_7)$ theory. Similarly to the $(A_2, A_6)$ case, the polynomials appearing in our affine scheme were determined, up to coefficients, from the associated vertex operator algebra in Section \ref{sec:higgsschemes}, and the coefficients are fixed via the matching with the Schur index. The affine scheme is
\begin{equation}
    X = \operatorname{Spec}(\mathbb{C}[x, y] / (x(2x^2 + 3y^3), y^2(5x^2 + y^3)) \,,
\end{equation}
where the variables $x$ and $y$ have the same grading as given in equation \eqref{eqn:A2A4grading}. Considering the $15$th jet scheme, we can again determine the leading orders of the Schur limit of the Hilbert series of the arc space using {\tt Macaulay2}. We find that the coefficients are
\begin{equation}
    (1, 0, 1, 2, 3, 4, 8, 10, 17, 23, 34, 47, 70, 92, 132, 177, 243, 322) \,,
\end{equation}
which matches the coefficients in the expansion of the Schur index of the $(A_2, A_7)$ theory as given in equation \eqref{eqn:schurAA}.

\subsection{\texorpdfstring{$(A_3, A_4)$}{(A3, A4)} theory}

We now consider the example of the $(A_3, A_4)$ Argyres--Douglas theory. Our affine scheme, as worked out following Section \ref{sec:higgsschemes}, is
\begin{equation}
    X = \operatorname{Spec}(\mathbb{C}[x, y, z] / (4xz + y^2 + 8z^3, xy + 4yz^2, x^2 + 4y^2z + 12xz^2 +  22z^4)) \,,
\end{equation}
where the coordinates have degrees 
\begin{equation}
    \operatorname{deg}(x) = 4 \,, \qquad \operatorname{deg}(y) = 3 \,, \qquad \operatorname{deg}(z) = 2 \,.
\end{equation}
Based on the degree of $z$ being $2$, we can use the jet scheme $J_{10}(R)$ to determine the coefficients of the Schur limit of the Hilbert series of the arc space up to and including order $q^{12}$. Again utilizing {\tt Macaulay2}, we find that the leading coefficients are
\begin{equation}
    (1, 0, 1, 2, 4, 5, 9, 12, 21, 29, 44, 60, 91) \,.
\end{equation}
As expected by this point, these coefficients match those in the expansion of the Schur index for the $(A_3, A_4)$ theory from equation \eqref{eqn:schurAA}.

\section{Macdonald index} \label{sec:mac}

In this section, we describe our prescription for computing the Macdonald index, a refined version of the Schur index. Our conjecture is that the Macdonald index is given by the Hilbert series of the associated graded algebra $\textrm{gr}(J_\infty(R))$ of the $J_\infty(R)$, with $X = \operatorname{Spec}(R)$. The prescription is essentially identical to that of \cite{Song:2016yfd}, except that we start with an arc space, instead of the full-fledged vertex operator algebra.\footnote{Essentially identical prescription to obtain the Macdonald grading from VOA is further studied in \cite{Bonetti:2018fqz, Agarwal:2018zqi, Beem:2019tfp, Xie:2019zlb, Watanabe:2019ssf, Agarwal:2021oyl}.} 
Let us describe this associated graded algebra.  

Let $V = J_\infty(R)$. Let us grade the monomials according to the ``degrees'', namely, by the number of letters we have: $\mathrm{gr}T(x_i) = 1$. 
If we have multiple generators $(x^{\ell})$ $\ell=1, 2, \cdots$, we can assign the grade as
$\mathrm{gr}T(x^{(\ell)}_i) = w(\ell)$ where $w$ can be different for each generator. For example, in the case of $W_N$-algebra, we assign $w(\ell) = \ell+1$ for $\ell=1, \cdots, N-1$ representing the currents of dimensions $2, 3, \cdots N$. 
Then we have a filtered algebra according to the grading we assigned to the generators
\begin{align}
    V_0 \subset V_1 \subset V_2 \subset \cdots \subset V \,, 
\end{align}
where
\begin{align}
    V_m = \{ v \in V : \mathrm{gr}T(v) \le m \} \,. 
\end{align}
From here, we can define a graded vector space by taking quotients as 
\begin{align} \label{eq:grV}
    \textrm{gr}V = \bigoplus_{m=0}^\infty \textrm{gr}V^{(m)} = \bigoplus_{m=0}^\infty (V_m/V_{m-1}) \ , 
\end{align}
where $V_{-1}$ is set to be trivial. 
From here, we can recover the Macdonald grading by evaluating the $T$-graded Hilbert series of $ \textrm{gr}V = \textrm{gr}(J_\infty(R))$:
\begin{align}
    I_{\textrm{Mac}} (q, T) = \textrm{HS}_{q, T} \left(\textrm{gr}(J_\infty(R)) \right) \equiv \sum_{m\ge 0} \textrm{HS}_q \left(\textrm{gr}(J_\infty(R))^{(m)} \right) T^m \,. 
\end{align}
Effectively, this way of introducing an extra grading is tantamount to counting the number of letters needed to construct a (basis) vector in $\textrm{gr}V$ endowed with grading $w$. When there is more than one type of letter, then the relation may not respect this grading based on the degree of letters. For example, when we have $x^2 = y^3$ with $w(x)=2$ and $w(y)=1$, the LHS has $w=4$ whereas RHS has $w=3$. The relation implies that we should not count both. 

Our prescription is simply that whenever the relation mixes the $T$-grading, drop the states with the highest $T$-weight. This is realized by equation \eqref{eq:grV}. For example, suppose the relation identifies the states with $T$-weight of $m$ and $m-1$. Then we see that such a vector of $T$-weight $m$ should be absent in $\textrm{gr}V^{(m)} = V_m/V_{m-1}$ from the relation. However such a vector is not removed in $\textrm{gr}V^{(m-1)} = V_{m-1}/V_{m-2}$ since the relation is in $V_m$ but not in $V_{m-1}$.  

This construction becomes trivial when there is only a single generator, such as the case of $(A_1, A_{2n})$ or a single family of generators having the same degree with a homogeneous relation. In such a case, the $T$-grading is already present at $V=J_\infty R$, so that $\textrm{gr}(V)$ is identical to $V$. 

\subsection{\texorpdfstring{$(A_2, A_3)$}{(A2, A3)} theory}

We begin by illustrating our procedure with the first non-trivial example. Consider the singularity defined by the affine scheme $X=\operatorname{Spec}(R)$, where
\begin{equation}
R = \mathbb{C}[x, y]/(x^2 - y^3, xy)\,.
\end{equation}
The arc space associated with $R$ is then given by 
\begin{align}
    J_\infty(R) = \mathbb{C}[x_0, y_0, x_1, y_1, \cdots]/( f_0, g_0, f_1, g_1, \cdots ) \,,
\end{align}
with
\begin{align}
    f_n = \sum_{i=0}^n x_i y_{n-i} \ , \qquad g_n = \sum_{i=0}^n x_i x_{n-i} - \sum_{i, j=0}^n y_i y_j y_{n-i-j} \,. 
\end{align}
Now, let us denote $V_{n, m_1, m_2}$ as the $\mathbb{C}$-vector space spanned by monomials with $m_1$ of $x_i$s, and $m_2$ if $y_i$s with the arc/jet degree adds up to $n$: 
\begin{equation}
    V_{n, m_1, m_2} \equiv \left\langle x_{i_1} \cdots x_{i_{m_1}} y_{j_1} \cdots y_{j_{m_2}} \,\,\bigg|\,\, \sum_{a=1}^{m_1} i_a + \sum_{b=1}^{m_2} j_b = n \right\rangle \,.
\end{equation}

\begin{itemize}
\item Level 1: For $V_{n, 1, 0}$ and $V_{n, 0, 1}$, generated by $x_n$ and $y_n$ respectively, no relations enter at this level. Thus, we get 
\begin{align}
    \sum_{n=0}^\infty V_{n, 1, 0} \to \frac{p_1}{1-q}\ , \quad \sum_{n=0}^\infty V_{n, 0, 1} \to \frac{p_2}{1-q} \,.
\end{align}

\item Level 2:
First we consider $V_{n, 0, 2}$ which is generated by $y_i y_{n-i}$. There are no relations at this level, and thus we find that\footnote{We use $(q)_n = (q; q)_n$ as a shorthand for the $q$-Pochhammer symbol.}
\begin{align}
  \sum_{n=0}^\infty V_{n, 0, 2} \to \frac{p_2^2}{(q)_2} \,. 
\end{align}
For $V_{n,1,1}$, there is one relation. We find that $xy=0$ removes a generator for each $n$. Hence we find that
\begin{align}
  \sum_{n=0}^\infty V_{n, 1, 1} \to \frac{p_1 p_2}{(1-q)^2} -\frac{p_1 p_2}{(1-q)} = \frac{q \, p_1 p_2}{(q)_1 (q)_1} \,. 
\end{align}
Finally, for $V_{n, 2, 0}$, generated by $x_i x_{n-i}$, the relation $x^2 = y^3$ effectively sets $\sum_{i=0}^n x_i x_{n-i} = 0$ at this level. It follows that
\begin{align}
    \sum_{n=0}^\infty V_{n, 2, 0} \to \frac{p_1^2}{(q)_2} - \frac{p_1^2}{(q)_1} = \frac{q^2 \, p_1^2 }{(q)_2} \,. 
\end{align}

\item Level 3: 
For $V_{n, 0, 3}$, which is generated by $y_i y_j y_{n-i-j}$, we find again that it is freely generated. The relation $x^2=y^3$ was already used to remove the states in $V_{n,2,0}$, which has a higher $T$-grade. Hence, we simply get
\begin{align}
    \sum_{n=0}^\infty V_{n, 0, 3} \to \frac{p_2^3}{(q)_3} \,. 
\end{align}
For the vector space $V_{n, 3, 0}$, generated by $x_i x_j x_{n-i-j}$, we have to consider the relation among the relations (i.e., the first syzygy). This is the same counting as we have already done for the $(A_1, A_2)$ case, since effectively we have $x(t)^2 = 0$. Hence we find
\begin{align}
    \sum_{n=0}^\infty V_{n, 3, 0} \to \frac{p_1^3}{(q)_3} - p_1^3\left(\frac{q}{(1-q)^2}+1 \right) = \frac{q^6 \, p_1^3}{(q)_3} \ . 
\end{align}
For $V_{n, 2, 1}$, we have checked by an explicit computation that the dimension is zero for $n=0, 1, 2$ and one for $n=3$. And for $V_{n, 1, 2}$, we also explicitly checked that the dimension is zero for $n=0, 1$ and one for $n=2$. 
\end{itemize}

Combining the above results, we find the leading orders of the Hilbert series of $\operatorname{gr} (J_\infty(R))$ to be  
\begin{align}
\begin{split}
    \operatorname{HS}_{p, q, T}(\operatorname{gr}(J_\infty(R))) &= 1 + \frac{p_1}{(q)_1} + \frac{p_2}{(q)_1} + \frac{q^2 \, p_1^2}{(q)_2} + \frac{q \, p_1 p_2}{(q)_1 (q)_1} + \frac{p_2^2}{(q)_2} \\
    &~~~ + \frac{q^6 \, p_1^3}{(q)_3} + \frac{q^3 \, p_1^2 p_2}{(q)_2 (q)_1} +  \frac{q^2 \, p_1 p_2^2}{(q)_1 (q)_2} + \frac{p_2^3}{(q)_3} + \frac{q^2 \, p_2^4}{(q)_4} + \cdots \,.
\end{split}
\end{align}
Here we have set $p_1 = p^3 T^2$, $p_2 = p^2 T$, which gives us a triply-graded Hilbert series (as a function of $p, q, T$) where $p$ grades the degree of the monomial, and $q$ grades the jet (physically identified with conformal descendant level) and $T$ is the Macdonald grading. To recover the Macdonald index, we set $p=q$ (or $p_1 = q^3 T^2$, $p_2 = q^2 T$):
\begin{align}
\begin{split}
    I_{\textrm{Mac}}(q, T) &= 1 + T q^2 + (T + T^2) q^3 + (T + 2 T^2) q^4 + (T + 2 T^2) q^5 \\
    &~~~ + (T + 3 T^2 + 2 T^3) q^6 
     + (T + 3 T^2 + 3 T^3) q^7 + (T + 4 T^2 + 5 T^3 + T^4) q^8 \\
    &~~~ + (T + 4 T^2 + 7 T^3 + 2 T^4) q^9 
    + (T + 5 T^2 + 9 T^3 + 5 T^4) q^{10} \\
    &~~~ + (T + 5 T^2 + 11 T^3 + 7 T^4 + T^5) q^{11} + O(q^{12}) \,.
\end{split}
\end{align}
To recover the Schur index, we further set $T = 1$. 

It seems our expression naturally gives rise to a new type of fermionic (or Nahm-sum) expression for the character of the $W(3, 7)$ minimal model, of the form:\footnote{In fact, this a refinement of the character of the minimal model by introducing a $T$-grading.}
\begin{align}
    \chi^{(\textrm{ref})}_{W(3, 7)} (q, T) = \operatorname{HS}_{q, q, T}(\operatorname{gr}(J_\infty(R))) = \sum_{k, \ell \ge 0} \frac{q^{f(k, \ell)} \, (q^3 T^2)^k (q^2 T)^\ell}{(q)_k (q)_{\ell}}
\end{align}
It would be interesting to find such a formula, which is useful for various applications, e.g., constructing the 3d TQFT whose boundary realizes the chiral $W$-algebra as in \cite{Gang:2023rei, Gang:2024loa}.

\section{Beyond points: Higgsable Argyres--Douglas theories} \label{sec:higgsable}

Thus far, we have considered our conjectures for 4d $\mathcal{N}=2$ SCFTs with a trivial Higgs branch. In this section, we study a small number of examples where the Higgs branch is non-trivial and show that our conjecture to obtain the Schur and Macdonald indices from the arc space of our affine scheme also holds in that context.

\subsection{\texorpdfstring{$\mathbb{C}^2/\mathbb{Z}_2$}{C2/Z2}}

The simplest non-trivial Higgs branch would be given by the Kleinian singularity $\mathbb{C}^2/\mathbb{Z}_2$. Many known $\mathcal{N}=2$ SCFTs have Higgs branch given by this singularity. For example, the Argyres--Douglas theory $(A_1, D_{2n+1}) = \mathcal{D}_{2n+1}(SU(2))$ has Higgs branch $\mathbb{C}^2/\mathbb{Z}_2$. Now, we can promote this variety $\mathbb{C}^2/\mathbb{Z}_2$ to a scheme (in a multitude of different ways), and then construct the corresponding arc space, determine the Schur or Macdonald limits of the Hilbert series of the arc space, and attempt to match to known 4d $\mathcal{N}=2$ SCFTs with Higgs branch $\mathbb{C}^2/\mathbb{Z}_2$.

In fact, we find an interesting wrinkle here. If we start by considering the spectrum of the ring 
\begin{align}
    R = \mathbb{C}[x, y, z] / ( x^2 + y^2 + z^2 ) \ , 
\end{align}
and compute the Hilbert series of its arc space, we find
\begin{align}
    \mathrm{HS}_{p, q} (J_\infty(R)) \Big|_{p \to qT} = 1 + 3 q T + q^2 (3 T + 6 T^2) + q^3 (3 T + 9 T^2 + 10 T^3) + \cdots \,. 
\end{align}
Here, we have chosen the degrees for $(x, y, z)$ to be all $1$ and set $p=qT$. Since the relation $x^2+y^2+z^2=0$ among the generators (which has the same degree) is homogeneous with definite degree, we do not have to consider the graded space $\mathrm{gr}(J_\infty R)$ separately, which will be simply identical to $J_\infty(R)$. 
Therefore, we can obtain the putative Macdonald index via choosing $p=qT$. 
However, this Hilbert series is unsuitable as a Macdonald (or Schur when $T\to 1$) index of a 4d SCFT or the vacuum character of a VOA.\footnote{It should be okay as a vertex algebra without a conformal vector, but any vertex algebra obtained from a 4d SCFT must have a conformal vector.} Specifically, the stress-tensor (which has to be present via Sugawara construction) is absent since there is no extra term of the form $\frac{q^2 T}{1-q}$, apart from the descendants of the current multiplet, contributing  $\frac{qT}{1-q}$ to the index. 

With these preliminaries in hand, let us now consider the affine scheme defined by an ideal of the following form
\begin{equation}\label{eqn:right}
    X = \operatorname{Spec} \left( \mathbb{C}[x, y, z, w] / ( w - (x^2 + y^2 + z^2), wx, wy, wz) \right) \,,
\end{equation}
where each of the coordinates has degrees:
\begin{equation}
    \operatorname{deg}(x) = \operatorname{deg}(y) = \operatorname{deg}(z) = 1 \,, \qquad \operatorname{deg}(w) = 2 \,.
\end{equation}
We can see that once $X$ is reduced, we obtain the $\mathbb{C}^2/\mathbb{Z}_2$ singularity as a variety. We can use {\tt Macaulay2} to determine the Schur limit of the Hilbert series of the $7$th jet scheme, which should overlap with the Schur limit of the Hilbert series of the arc space up to and including the $q^8$ term. We find:
\begin{equation}
    \operatorname{HS}_{q,q}(J_\infty(R)) = 1 + 3q + 9q^2 + 19q^3 + 42q^4 +  81q^5 + 155q^6 + 276q^7 + 486q^8 + \cdots \,.
\end{equation}
To the leading orders, this expression agrees with the known Schur index of $\mathcal{D}_3(SU(2)) = (A_1, A_3) = (A_1, D_3)$ theory, which is
\begin{equation}
     \operatorname{PE}\left[ \frac{q - q^3}{(1-q)(1-q^3)} \chi_3  \right] \,.
\end{equation}
We see that these match perfectly at least up to and including order $q^8$. While it would be useful to go to higher orders to provide a more convincing demonstration of our conjecture in equation \eqref{eqn:schurconj}, {\tt Macaulay2} has difficulty finding Gr\"obner bases for the higher jet schemes. 
In the recent work \cite{Andrews:2025krn}, the authors proved that the arc space of our affine scheme indeed gives the Schur and Macdonald indices. 

In this analysis, we have chosen to introduce a degree-two coordinate $w$ corresponding to the stress-tensor of the 4d SQFT for convenience, however, the scheme $X$ is isomorphic to the scheme
\begin{equation}\label{eqn:wrongish}
    \operatorname{Spec} \left( \mathbb{C}[x, y, z] / ( (x^2 + y^2 + z^2)x, (x^2 + y^2 + z^2)y, (x^2 + y^2 + z^2)z) \right) \,,
\end{equation} 
where the indeterminates have the same degrees as before. It is necessary (see \cite{Kang:2026nge} for more details) for our affine scheme to be bifiltered if it is to reproduce the Macdonald index. We can specify the bifiltration in the ring in equation \eqref{eqn:right} by inheriting it from the bigrading
\begin{equation}
    \operatorname{deg}(x) = \operatorname{deg}(y) = \operatorname{deg}(z) = (1, 1) \,, \qquad \operatorname{deg}(w) = (2, 1) \,,
\end{equation}
on $\mathbb{C}[x,y,z,w]$. We can then push this bifiltration through the isomorphism to the ring in equation \eqref{eqn:wrongish}, however, this bifiltration cannot be inherited from a bigrading structure on the ring $\mathbb{C}[x,y,z]$. Therefore, we choose to include a coordinate $w$ corresponding to the stress-tensor explicitly to make the choice of bifiltration clear.

We can now consider a different, but similar, scheme that reduces to $\mathbb{C}^2/\mathbb{Z}_2$ as a variety. We now define $X$ via
\begin{align}
    X = \operatorname{Spec}\left( \mathbb{C}[x, y, z, w] / (w - (x^2 + y^2 + z^2), x w^2, y w^2, z w^2) \right) \,.
\end{align}
We find that the Schur limit of the Hilbert series of the arc space is
\begin{align}
    \operatorname{HS}_{q, q}(J_\infty(R)) &= 1 + 3q + 9q^2 + 22q^3 + 51 q^4 + 105 q^5 \\&\qquad\qquad + 212 q^6 + 402 q^7  + 744q^8 + 1326q^9 + 2316q^{10} +  \cdots \,, 
\end{align}
which agrees with the Schur index of the $(A_1, D_5) = \mathcal{D}_5(SU(2))$ theory, that is given as
\begin{align}
    I_{(A_1, D_5)}(q) = \operatorname{PE}\left[ \frac{q - q^5}{(1-q)(1-q^5)} \chi_3  \right] \,,
\end{align}
up to and including the $q^{10}$ term. Therefore, we conjecture that the following affine scheme captures the Macdonald and Schur indices of the $(A_1, D_{2n+1}) = \mathcal{D}_{2n+1}(SU(2))$ theory:
\begin{align}
\begin{split}
    X &= \operatorname{Spec}\left( \mathbb{C}[x, y, z, w] / (w - (x^2 + y^2 + z^2), x w^n, y w^n, z w^n) \right) \,.
\end{split}
\end{align}

There exist many 4d $\mathcal{N}=2$ SCFTs with identical Higgs branches, and there are infinitely many ways to ``scheme-ify'' the $\mathbb{C}^2/\mathbb{Z}_2$ singularity. For the current example, we distinguish them by the vanishing condition of the form $J_{i} \times T^n \sim 0$ ($J_i$ and $T$ denote current and stress-tensor, respectively), which gives rise to different 4d SCFTs. We naturally expect that there exist ``higher-dimensional'' versions of such constructions, giving rise to more 4d SCFTs. It would be interesting to see if it is possible to obtain all $\mathcal{N}=2$ SCFTs with the same Higgs branch moduli space starting by considering distinct affine schemes. 

\subsection{\texorpdfstring{$\mathbb{C}^2/\mathbb{Z}_{n+1}$}{C2/Zn+1}}

Now, let us consider 4d $\mathcal{N}=2$ SCFTs whose Higgs branches are the Kleinian singularities $\mathbb{C}^2/\mathbb{Z}_{n+1}$. 
As before, if we start with the simplest scheme which reduces to $\mathbb{C}^2/\mathbb{Z}_3$ as a variety with no nilpotents:
\begin{align}
    X = \operatorname{Spec} \left(\mathbb{C}[x, y, z]/(xy+z^3) \right) \,, 
\end{align}
then the Hilbert series of the corresponding arc space does not have the requisite properties to be the index of an interacting 4d $\mathcal{N}=2$ SCFT. Instead, we feed in an extra element, which will be mapped to the stress tensor. To this end, we can consider
\begin{align}\label{eqn:swiper}
   X = \operatorname{Spec} \left( \mathbb{C}[x, y, z, t]/( xy+z^3 - zt, xt, yt, z(xy+z^3) ) \right) \,. 
\end{align}
We assign degrees for $(x, y, z, t)$ to be $(3/2, 3/2, 1, 2)$. 
The Schur limit of the Hilbert series of the arc space of $R=\mathbb{C}[X]$ is found to be
\begin{align}
\begin{split}
   \operatorname{HS}_{q, q}(J_\infty(R)) &= 1 + q + 2q^{3/2} + 3q^2 + 4q^{3/2} + 7q^2 + 8q^{5/2} + 14 q^3 + 18 q^{7/2} \\
   &\qquad + 26 q^4 + 34q^{9/2} + 49 q^5 + 62 q^{11/2} + 86q^6 + \cdots \,.
\end{split}
\end{align}
Surprisingly, this result agrees with the known Schur index of the $(A_1, A_5)$ Argyres--Douglas theory \cite{Song:2017oew}
\begin{align}
    I_{(A_1, A_5)} = \operatorname{PE}\left[ \frac{q+(z^2+z^{-2})q^{3/2}+q^2 - q^3 - (z^2+z^{-2})q^{7/2} - q^4}{(1-q)(1-q^4)}\right] \,,
\end{align}
which is known to have $\mathbb{C}^2/\mathbb{Z}_3$ as its Higgs branch. The associated vertex operator algebra for this theory is given by the Bershadsky--Polyakov (BP) W-algebra \cite{Bershadsky:1990bg, Polyakov:1989dm}, whose connection to Argyres--Douglas theories was found in  \cite{Creutzig:2017qyf}. The BP algebra has strong generators of dimensions $1, 3/2, 3/2, 2$, and its Zhu's algebra $R_V$ is identical to our $R$ so that the associated scheme $X_V$ is identical to our affine scheme $X$.\footnote{There is a conjectural form of the Zhu's algebra written in \cite{Xie:2019zlb} for the $(A_1, A_{2n+1})$ theory, which is slightly different from what we have. This gives a different Hilbert series which does not match with the Schur index.} 

\section{Discussion}\label{sec:conc}

In this paper, we have proposed that we can associate an affine scheme $X$ to a 4d $\mathcal{N}=2$ SCFT, whose Schur index is then given by the appropriately-graded Hilbert series of the arc space of the polynomial ring $J_\infty(R)$ with $R=\mathbb{C}[X]$, thereby giving an algebro-geometric interpretation of the Schur index. One can also obtain the Macdonald index by considering a graded space $\textrm{gr}(J_\infty(R))$, giving rise to additional gradings, and by computing its Hilbert series. The reduced scheme of $X$ gives the Higgs branch moduli space of the 4d theory. 

We primarily focus on the case where $X$ is a point upon reduction to an affine variety. We find that different choices of $X$, encoding the singular behavior of a point in various dimensions, give rise to distinct Argyres--Douglas theories without a Higgs branch, such as $(A_{k-1}, A_{n-1})$ with $\gcd(k, n)=1$. The nilpotent elements in the affine scheme $X$ encode the vanishing of certain short multiplets in the 4d theory. When the reduced $X$ is non-empty, we need to make sure to include a coordinate that plays the role of a stress-energy tensor. One very interesting and surprising feature is that this simple nilpotency relation (such as $x^n=0$ or $x^2-y^3=xy=0$) is enough to determine the entire Schur or Macdonald index. It is easy to read off the leading relation when the index is given, but the index contains far more than just the leading relation. It is rather remarkable that a small number of ``primitive'' or ``defining'' relations is enough to determine (infinitely) many relations or also relations of relations. It would be interesting to understand the physical mechanism behind this. Also, one further interesting aspect of our construction is that we naturally obtain a ``fermionic'' formula for the Schur index or the character of the associated VOA. 

Our affine scheme $X$ is often (but not always) identical to the associated scheme $X_V$ of the vertex operator algebra $V$ for the 4d $\mathcal{N}=2$ SCFT. In this case, the arc space $J_\infty(R) = J_\infty(\mathbb{C}[X])$ of the polynomial ring over $X$ turns out to be identical to the arc space over Zhu's $C_2$-algebra $R_V$, which can be constructed from a VOA $V$ by throwing away higher derivatives. Compared to the full-fledged vertex operator algebra $V$, the arc space of $\mathbb{C}[X]$ is a much simpler object, and it does not contain local information about the OPE of Schur operators. So given a 4d theory $\mathcal{T}$, we have a vertex operator algebra $V_\mathcal{T}$, from which we obtain the commutative algebra $R_{V_\mathcal{T}}$ and the affine scheme $X_{V_\mathcal{T}}$:
\begin{align}
    \mathcal{T} \leadsto V_{\mathcal{T}} \leadsto R_{V_\mathcal{T}} \leadsto X_{V_\mathcal{T}} \,.
\end{align}
On the other hand, our affine scheme can reproduce the Schur and Macdonald indices, as we have discussed. This hints that it may be possible to reconstruct the entire VOA (including the central charge) from the affine scheme $X$ and passing to the arc space of $\mathbb{C}[X]$ and eventually the full 4d (relative) theory (modulo moving on the conformal manifold):
\begin{align}
    X \leadsto J_\infty(\mathbb{C}[X]) \overset{\mathclap{?}}{\leadsto} V_X \overset{\mathclap{?}}{\leadsto} \mathcal{T}_X\,.
\end{align}
We have demonstrated that, for several examples, it is indeed possible to reconstruct Schur and Macdonald indices; however, it is not even obvious whether $V_X$ is uniquely determined by $X$. We have already seen that we may get different characters if we choose $p \neq q$. For example, the double point case $\mathbb{C}[x]/(x^2)$ gave the vacuum character of the $M(2, 5)$ minimal model or $\mathfrak{osp}(1|2)_1$ depending on the specialization of the fugacities in the Hilbert series. The latter can also be thought of as a non-vacuum character of $M(2, 5)$. Therefore, it is feasible that we get many VOAs from the same $X$ via taking different limits. However, if we demand a non-unitary VOA of \cite{Beem:2013sza}, it may be possible to uniquely fix the associated VOA. 

As we have seen, not all affine schemes give rise, via the arc space construction, to indices consistent with coming from a 4d SCFT. Because of $\mathcal{N}=2$ supersymmetry, we expect that $X$ upon reduction should be described by a certain hyperk\"ahler cone/symplectic singularity. But even with fixed $X_\textrm{red}$, not all unreduced schemes $X$ qualify to give a good SCFT index. It would be interesting to find a geometric criterion for $X$ such that it realizes the indices of a 4d SCFT. A concept analogous to quasi-lisse VOA might be useful \cite{arakawa2018quasi}. 

Let us conclude with a few possible future research directions: 
\begin{itemize}
    \item More examples from a point: it would be useful to consider other point-like singularities, which will likely produce other higher-rank Argyres--Douglas theories without a Higgs branch. In particular, we have shown that every scheme over a point defined via the quotient of polynomial ring in a single variable, $\mathbb{C}[x]$, with a single homogeneous polynomial generating the quotient ideal can be associated to a unique Argyres--Douglas theory. A next step could be to determine all schemes over a point in quotients of two-variable polynomial rings, and ask which such affine schemes can be associated with the indices of 4d $\mathcal{N}=2$ SCFTs with no Higgs branch.
    \item Higgs branch RG flows and non-Higgsable sectors: each $\mathcal{D}_p(SU(2))$ Argyres--Douglas theory with $p$ odd has Higgs branch isomorphic to $\mathbb{C}^2/\mathbb{Z}_2$, however at the generic point of the Higgs branch there exists a non-Higgsable SCFT (that is, a theory where the Higgs branch is a point), which is the $(A_1, A_{p-3})$ theory. This non-Higgsable sector is opaque to the Higgs branch and the Higgs branch chiral ring, however, our affine scheme, which is a scheme-ification of the Higgs branch, is sensitive; we observe that our affine scheme encodes these non-Higgsable structures. Furthermore, the Higgs branch is a symplectic singularity, and the foliation structure of the symplectic singularity captures all Higgs branch renormalization group flows \cite{Bourget:2019aer}; does there exist an analogous foliation structure on our affine scheme that captures information of RG flows beyond the Higgs branch as a variety?
    \item More examples with non-empty Higgs branch: it would be interesting to work out more examples, especially other Kleinian singularities or nilpotent orbits. There are many 4d SCFTs whose Higgs branches are given by such varieties. For example: the Argyres--Douglas theory $\mathcal{D}_2(SU(3))$ has a Higgs branch which is the closure of the minimal nilpotent orbit of $\mathfrak{su}(3)$; Zhu's $C_2$-algebra $R_V$ was given explicitly in \cite{Xie:2019zlb}, and thus working out the Hilbert series of the arc space is straightforward, though computationally challenging. Some such 4d theories include class $\mathcal{S}$ theories \cite{Gaiotto:2009hg, Gaiotto:2009we}. Theories with higher supersymmetry are likely to be interesting. 
    \item A common source of pathologies in our understanding of 4d $\mathcal{N}=2$ SCFTs occurs when we consider theories that arise via the gauging of a discrete global symmetry of another theory \cite{Argyres:2017tmj,Argyres:2018wxu,Bourget:2018ond,Bourton:2018jwb}. A good stress-test of our proposal would be to determine if there exists an affine scheme that captures the indices for simple examples of discretely-gauged 4d $\mathcal{N}=2$ SCFTs. The discrete-gauging often acts on the Higgs branch operators in a clear way, see, e.g., \cite{Giacomelli:2024sex,Lawrie:2025exx}, and it will be key to understand how this action uplifts to the non-reduced affine scheme $X$.
    \item Thus far we have focused on relative \cite{Freed:2012bs} 4d $\mathcal{N}=2$ SCFTs; to fully-specify the QFT it is necessary to also give a choice of polarization of intermediate defect group \cite{Belov:2006jd}.\footnote{In fact, it is necessary to specify a polarization \emph{pair} to incorporate possible SPT phases of the vacuum of the theory \cite{Lawrie:2023tdz}.} For gauge theory this is related to a choice of global form of the gauge group for a fixed gauge algebra. The VOA is known to encode the Lens space index \cite{Fluder:2017oxm}, which is sensitive to this global structure of the 4d theory. Therefore, we would ask if our affine scheme $X$, or a refinement, also encodes this choice of global structure.
    \item So far, our affine scheme $X$ is an auxiliary object of the 4d theory, and the moduli space of vacua appears only after reduction; such moduli spaces can often be realized via geometric engineering in string/M-theory. It is natural to ask whether we can recover the unreduced $X$, with the full nilpotent structure, directly via (algebro-)geometric engineering in string theory. 
\end{itemize}
We intend to address some of these questions in future work.

\subsection*{Acknowledgements}
We thank Tomoyuki Arakawa, Heeyeon Kim, Lorenzo Mansi, and Tomas Prochazka for useful discussions. We also thank Jacques Distler for comments on a previous version of this paper. We thank the Simons Center for Geometry and Physics and the organizers of the Supersymmetric Quantum Field Theories, Vertex Operator Algebras, and Geometry program, where this work was conceived. 
M.~J.~K.~is supported by the U.S.~Department of Energy, Office of Science, Office of High Energy Physics, under Award Numbers DE-SC0013528 and QuantISED Award DE-SC0020360.
M.~J.~K.~is also supported by the Start-up Research Grant for new faculty provided by Texas A\&M University.
C.L.~acknowledges support from DESY (Hamburg, Germany), a member of the Helmholtz Association HGF; C.L.~also acknowledges the Deutsche Forschungsgemeinschaft under Germany's Excellence Strategy - EXC 2121 ``Quantum Universe'' - 390833306 and the Collaborative Research Center - SFB 1624 ``Higher Structures, Moduli Spaces, and Integrability'' - 506632645.
The work of J.S.~is supported by the National Research Foundation of Korea (NRF) grants RS-2023-00208602 and RS-2024-00405629, and also by the KAIST-KIAS collaboration program. 

\bibliography{references}{}
\bibliographystyle{sortedbutpretty}
\end{document}